# A hybrid pulsed laser deposition approach to grow thin films of chalcogenides

Mythili Surendran[1,2], Shantanu Singh[1], Huandong Chen[1], Claire Wu[1], Amir Avishai[2], Yu-Tsun Shao[1,2], and Jayakanth Ravichandran[1,2,3]*

[1]Mork Family Department of Chemical Engineering and Materials Science, University of Southern California, Los Angeles, California 90089, USA
[2]Core Center for Excellence in Nano Imaging, University of Southern California, Los Angeles, California 90089, USA
[3]Ming Hsieh Department of Electrical and Computer Engineering, University of Southern California, Los Angeles, California 90089, USA

*e-mail: j.ravichandran@usc.edu

**Abstract**

Vapor-pressure mismatched materials such as transition metal chalcogenides have emerged as electronic, photonic, and quantum materials with scientific and technological importance. However, epitaxial growth of vapor-pressure mismatched materials are challenging due to differences in the reactivity, sticking coefficient, and surface adatom mobility of the mismatched species constituting the material, especially sulfur containing compounds. Here, we report a novel approach to grow chalcogenides - hybrid pulsed laser deposition - wherein an organosulfur precursor is used as a sulfur source in conjunction with pulsed laser deposition to regulate the stoichiometry of the deposited films. Epitaxial or textured thin films of sulfides with variety of structure and chemistry such as alkaline metal chalcogenides, main group chalcogenides, transition metal chalcogenides and chalcogenide perovskites are demonstrated, and structural characterization reveal improvement in thin film crystallinity, and surface and interface roughness compared to the state-of-the-art. The growth method can be broadened to other vapor-pressure mismatched chalcogenides such as selenides and tellurides. Our work opens up opportunities for broader epitaxial growth of chalcogenides, especially sulfide-based thin film technological applications.

Emerging electronic, photonic, and quantum materials such as transition metal chalcogenides have demonstrated novel physical phenomena that are being harnessed for device applications[1-5]. These materials tend to be vapor pressure mismatched, *i.e.,* the constituent components of the material have significantly different vapor pressure for the corresponding precursors. This mismatch results in large differences in reactivity, sticking coefficient, and surface adatom mobility, which in turn creates challenges in phase formation, and crystallization of these materials. For example, early transition metals such as Zr and Hf have significantly different vapor pressure compared to sulfur, which makes the growth of potential high mobility semiconductors such as $ZrS_2$, $HfS_2$, an outstanding synthesis challenge. This challenge is compounded, when one considers ternary/complex stoichiometries such as chalcogenide perovskites $BaZrS_3$ and their layered counterparts, Ruddlesden Popper phases such as $Ba_3Zr_2S_7$. Specifically, regulating the anionic stoichiometry in these materials is challenging due to competing deleterious reactions such as oxidation. Hence, there is a need to develop advanced synthesis techniques to overcome the thermodynamic and kinetic barriers for phase formation, composition, and crystallinity during thin film growth of vapor pressure mismatched materials.

Thin film growth of vapor pressure mismatched materials such as transition metal chalcogenides (TMCs) and transition metal pnictides is in their infancy due to the challenges in the synthesis of these materials[6-8]. Among chalcogenides (unlike selenides and tellurides[9, 10]), epitaxial growth of sulfides is poorly explored, due to the mismatched low vapor pressure cations and high vapor pressure sulfur, low sticking coefficient of sulfur[11, 12], and difficulties in cracking sulfur and $H_2S$[13-16], the two most common precursors for sulfides. The challenges become acute for TMCs, especially complex ternary or quaternary TMCs. Although high temperature chemical vapor deposition (CVD) is widely employed to grow large area TMCs due to its simplicity, the development of other vapor phase techniques such as MBE and PLD are desired for improved structural and chemical control over the materials at significantly lower growth temperatures. This is especially important due to the recent interest in chalcogenides of various forms (*e.g.,* 2D transition metal dichalcogenides) investigated heavily for their electronic and optoelectronic properties. In the recent years, chalcogenide perovskites such as $BaZrS_3$ and its Ruddlesden Popper phases have also gained substantial interest due to suitable band gap and high absorption coefficient for next-generation photovoltaics[17-21]. Meanwhile, quasi-1D hexagonal perovskites such as $BaTiS_3$ and $Sr_{1+x}TiS_3$ exhibit large optical anisotropy in the infrared region and host charge density wave-like electronic phase transitions[22-24]. Thin film growth of these materials is expected to find broad electrical and photonic applications, especially as the breadth of semiconductors desired in the CMOS platform has grown considerably recently.

Sulfur is extremely volatile at high temperatures and a sulfur over pressure is needed to prevent desorption and facilitate stoichiometric growth. Most of the existing growth techniques rely on H$_2$S, a toxic, hazardous, and flammable gas with poor decomposition efficiency (pure or diluted in argon gas) as a sulfur source. Elemental sulfur powder is a more benign replacement for H$_2$S, but the decomposition of sulfur is very complicated and inefficient yielding a mixture of S$_n$ allotropes ($n$ = 2-10) depending on the temperature[16, 25]. This is a major issue for vapor phase thin film deposition techniques such as MBE and PLD, where sulfur contamination and constant degassing of sulfur affects the vacuum and reliability of the growth systems. Moreover, transition metal sulfides require excellent cationic and anionic stoichiometry control as they can stabilize in a wide range of poly-sulfides and modulated structures. Thus, the developed growth methods will need to pay particular attention to the precursors that regulate the anionic stoichiometry.

In this work, we report a novel hybrid pulsed laser deposition (PLD) approach for the growth of sulfides, motivated by the hybrid MBE and MOCVD growth strategies.[26, 27] In the hybrid PLD, we employ an organosulfur precursor as an alternative to sulfur precursors such as H$_2$S. This method shows an efficient incorporation of sulfur during PLD, as the sulfur precursors decompose at much lower temperatures (~250-400°C) with significantly higher efficiency compared to other sources such as elemental sulfur and H$_2$S. The availability of high concentration of active sulfur species at relatively low temperatures is beneficial to maintain the sulfur stoichiometry at sufficiently high vacuum conditions, while the cationic stoichiometry is regulated by the target. We report epitaxial or textured growth of a broad array of binary and complex sulfides such as alkaline-earth binary sulfides, transition metal disulfides and perovskite sulfides, demonstrating the versatility of this method. This approach can be extended to other chalcogenides such as selenides and tellurides, wherein a suitable organo-precursor shall be employed as chalcogen precursor. Further, metal-organic precursors can be easily added to this approach for compensating volatile cationic species and/or doping studies.

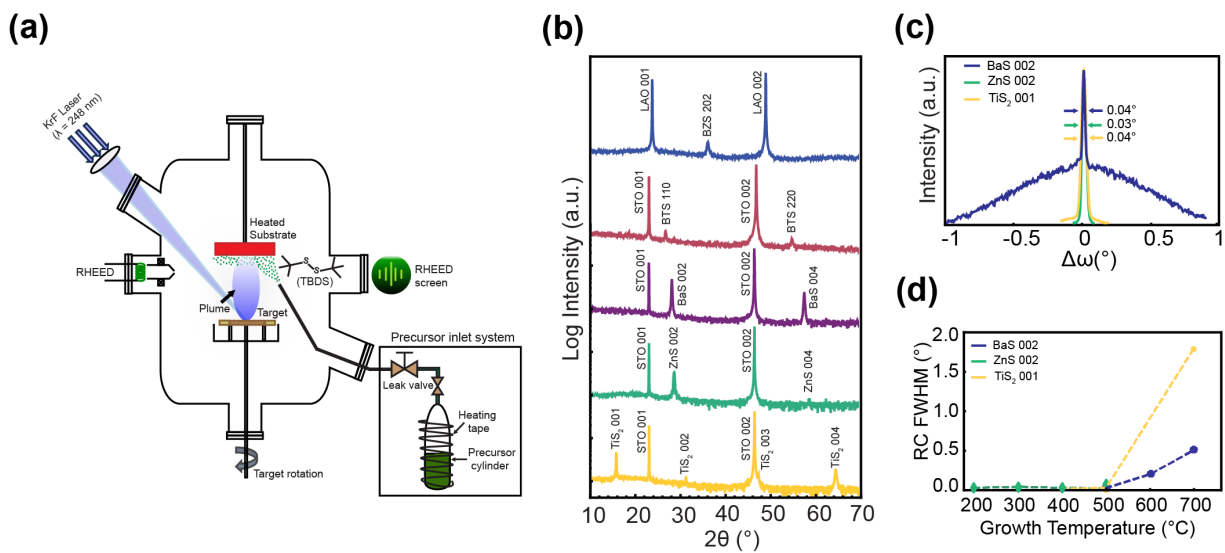

**Fig. 1(a)** Schematic of hybrid pulsed laser deposition showing the organosulfur precursor delivery system. **(b)** High resolution XRD pattern of a representative BZS film on LAO substrate, and BTS, BaS, ZnS and TiS$_2$ films on STO substrates indicating strong out-of-plane texture. **(c)** Rocking curve (RC) FWHM for BaS and TiS$_2$ films grown at 500°C and ZnS film grown at 400°C on STO. **(d)** RC FWHM as a function of growth temperature for BaS, ZnS and TiS$_2$ thin films.

## Results and Discussion

Fig.1a) shows the schematic illustrating the hybrid PLD process for the growth of chalcogenides. The design considerations for this novel method are discussed in Supplementary information. We grew thin films of structurally and chemically diverse sulfides such as alkaline earth binary sulfides (3D) and TMCs (layered/2D), and ternary or complex sulfides such as chalcogenide perovskites (3D and 1D) using the hybrid PLD approach. In Fig. 1b), we show representative XRD patterns of the thin films of sulfides such as BaZrS$_3$ (BZS), BaTiS$_3$ (BTS), BaS, ZnS, and TiS$_2$ with strong out-of-plane texture. BZS thin films were grown on LaAlO$_3$ substrates, while all the other sulfides namely BTS, BaS, TiS$_2$ and ZnS were grown on SrTiO$_3$ (STO) substrates. The rocking curves of BaS 002, ZnS 002 and TiS$_2$ 001 are shown in Fig. 1c) with a narrow full width at half maximum (FWHM) of ≤0.04° suggesting high crystalline quality with minimal incoherencies or misorientation of grains. The rocking curve FWHMs remain low up to growth temperatures of about 500°C (Fig. 1d)) and increase significantly at higher temperatures. This is presumably due to the sulfur vacancies or other interfacial defects associated with interdiffusion and reactivity at high growth temperatures.

BZS and BTS shown in figure 1b were grown at 750°C and 700°C respectively and their crystallinity was similar to those grown by conventional PLD in argon – H$_2$S (5%) as previously reported by our group.[28, 29] However, the surface and interface smoothness was significantly better, which will be discussed later. This demonstrates the efficiency of TBDS as a sulfur precursor alternative to H$_2$S for the growth of complex chalcogenides. We observed only BZS 202 and BTS 110 family of reflections in the out-of-plane scans indicating a strong preferred orientation in these films. The off-axis $\phi$-scans and grazing incidence small angle X-ray scattering (GIWAXS) show the epitaxial and quasi-epitaxial nature of the films respectively, similar to previous studies. However, binary sulfides such as BaS, SrS, CaS, TiS$_2$ and ZnS were successfully grown at BEOL compatible temperatures (450°C or below). Alkaline earth binary chalcogenides (such as BaS, SrS and CaS) crystallize in a rock salt structure with space group $Fm\bar{3}m$. The films were fully relaxed and showed strong *c*-axis oriented texture (00*l* reflections in the XRD scans in Fig. 1b and Supplementary Fig. S2) and smooth surfaces and interfaces (X-ray reflectivity (XRR) and AFM topography in Supplementary Fig. S3). The rocking curve FWHM of 002 peak of BaS thin film grown at 500°C was 0.03° (figure 1c), with a broader background indicating incoherencies arising from large in-plane misorientations of the grains. At higher temperatures, the rocking curve FWHM significantly increased (figure 1d) presumably due to the interfacial damage and reactivity. This aligns with the larger rocking curve values reported by previous studies on epitaxial thin films of doped SrS on MgO substrate grown at 750-800°C using H$_2$S[30, 31]. However, the films grown using hybrid PLD were highly crystalline and remained epitaxial at temperatures as low as 400°C, even at large film-substrate misfits.

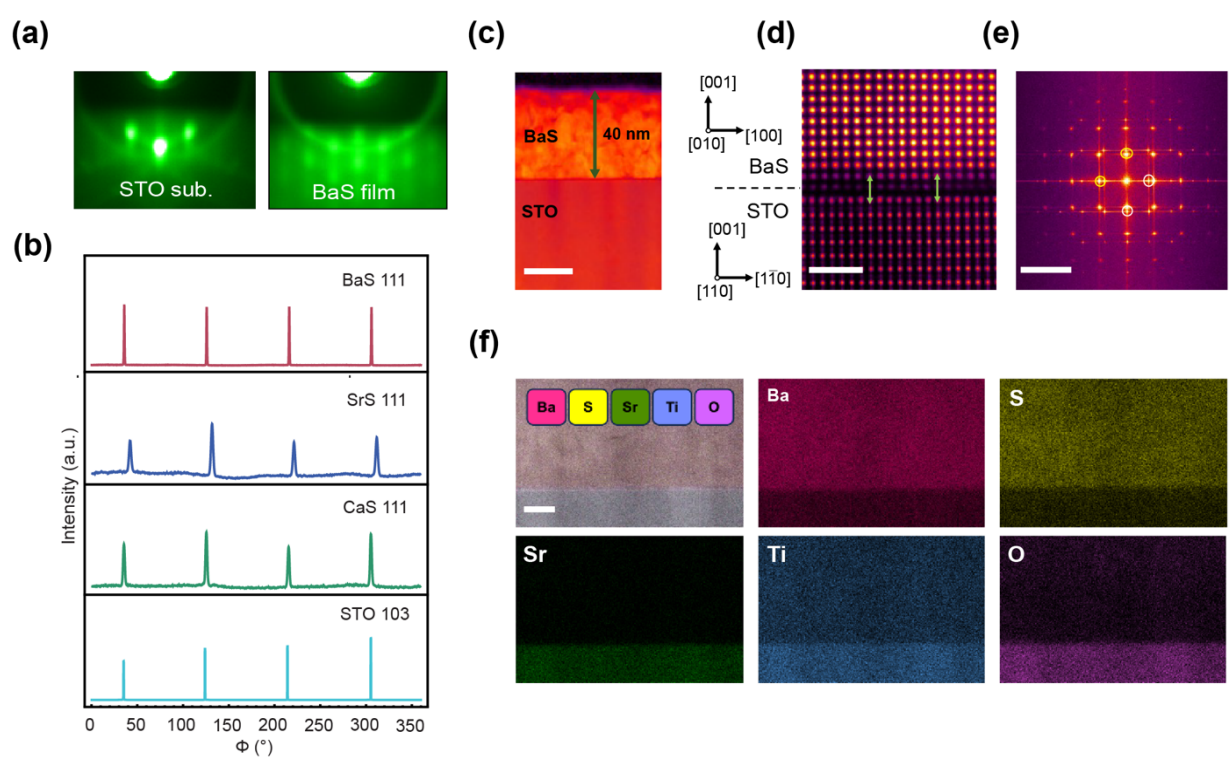
5

**Fig. 2**. Representative RHEED pattern for **(a)** annealed STO substrate prior to deposition and BaS thin film after deposition showing 3D-streaky pattern. **(b)** Off-axis $\phi$ scans of BaS 111, SrS 111, CaS 111and STO 103 showing epitaxial relationship. **(c)** Wide field-of-view HAADF image showing 40 nm BaS. **(d)** Atomic resolution HAADF image for BaS/STO interface indicating the epitaxial orientation of the film with respect to the substrate. **(e)** FFT patterns for the region in (d) showing BaS 100/001 diffraction spots (yellow circles) along [110] zone axis of STO. **(f)** Survey HAADF image and corresponding elemental maps for Ba, S, Sr, Ti and O showing uniform distribution of elements. Scale bars correspond to 20 nm for c), 2 nm for (d) 5 nm$^{-1}$ for (e) and 10 nm for (f).

As shown in Fig. 2a, a streaky RHEED pattern was observed during the BaS growth (along [100] direction of annealed STO substrate) suggesting a definite in-plane orientation for BaS films. The epitaxial relationship was determined by off-axis $\phi$-scans along BaS (SrS, CaS) 111 and STO 103 and observed four peaks each for BaS 111, SrS 111, CaS 111 and STO 103 separated by 90 degrees and aligned to each other as shown in Fig. 2b. This implies that the direct epitaxial relationship between the 45° rotated film lattice along [110] direction and the [100] direction of the substrate is (001) [110] BaS (SrS, CaS) // (001) [100] STO. To confirm the epitaxy and investigate the interfacial quality of hybrid PLD grown BaS thin films, we performed scanning transmission electron microscopy (STEM) studies. Fig. 2c shows a representative wide field-of-view high resolution high-angle annular dark field (HAADF) image of ~40 nm thick BaS film on STO substrate. The epitaxial relationship was confirmed as shown in Fig. 2d. Similar epitaxial relationship exists for SrS and CaS films grown on STO (Supplementary Fig. S4). However, the films were polycrystalline with extended defects such as dislocations and grain boundaries due to the large lattice mismatch between the films and the STO substrate. Interestingly, BaS/STO interface had better crystalline quality compared to SrS and CaS, despite having the largest lattice mismatch. This can be clearly seen from the XRD rocking curve FWHM measurements as well. We speculate the existence of domain matching epitaxy at the interface of BaS and STO resulting in a lower misfit of about 1.1%. Such a lattice matching was observed in local regions of the BaS/STO interface where dislocations exist at a period of seven lattice planes of BaS along [100] and eight lattice planes of STO along [110] directions (green arrows in Fig. 2d). The FFT pattern in figure 2e indicate diffraction spots corresponding to both BaS (yellow circle) and STO (white circle) indicating definite in-plane orientation. Fig. 2f shows a HAADF image where the high-resolution energy-dispersive X-ray spectroscopy (EDS) spectra were acquired and the corresponding elemental maps of Ba, S, Sr, Ti and O reveal a uniform chemical composition of the film. The film-substrate interface shows no evident intermixing of elements across the interface.

In addition to alkaline earth chalcogenides, we also grew ZnS (3D main group chalcogenide) and TiS$_2$ (2D TMDC) films STO substrate with strong out-of-plane texture as evidenced by the ZnS 00$l$ and TiS$_2$ 00$l$ reflections in the XRD scan (Fig. 1b). In case of ZnS, the film surface and interface were extremely smooth as indicated by the Pendellösung fringes in the XRD pattern and the Kiessig fringes in XRR (Supplementary Fig. S5a&b). This was further confirmed by AFM topography scan indicating an RMS roughness of about 0.35nm (Supplementary Fig. S5b inset). In this work, the rocking curve FWHM of 002 peak of ZnS grown at 400°C was found to be 0.03° (Figure 1c) showing a superior crystalline quality and is the lowest reported thus far by various growth techniques[32-34]. ZnS films were crystallized at temperatures as low as 200°C (Supplementary Fig. S6) although the intensity of ZnS 002 peak increased with temperature while the rocking FWHM did not change significantly. This shows that the crystallization of ZnS is not limited by decomposition of TBDS, and it follows the expected trend of increased crystallinity with temperature, while TBDS ensures the availability of decomposed sulfur at higher temperatures for stoichiometric composition. This can be seen from the fact that ZnS 002 peak shifts to a lower *2θ* value (higher inter-planar spacing) under vacuum growth conditions (no TBDS used) at 400°C indicating presence of point defects such as sulfur vacancies in the film (Supplementary Fig. S7).

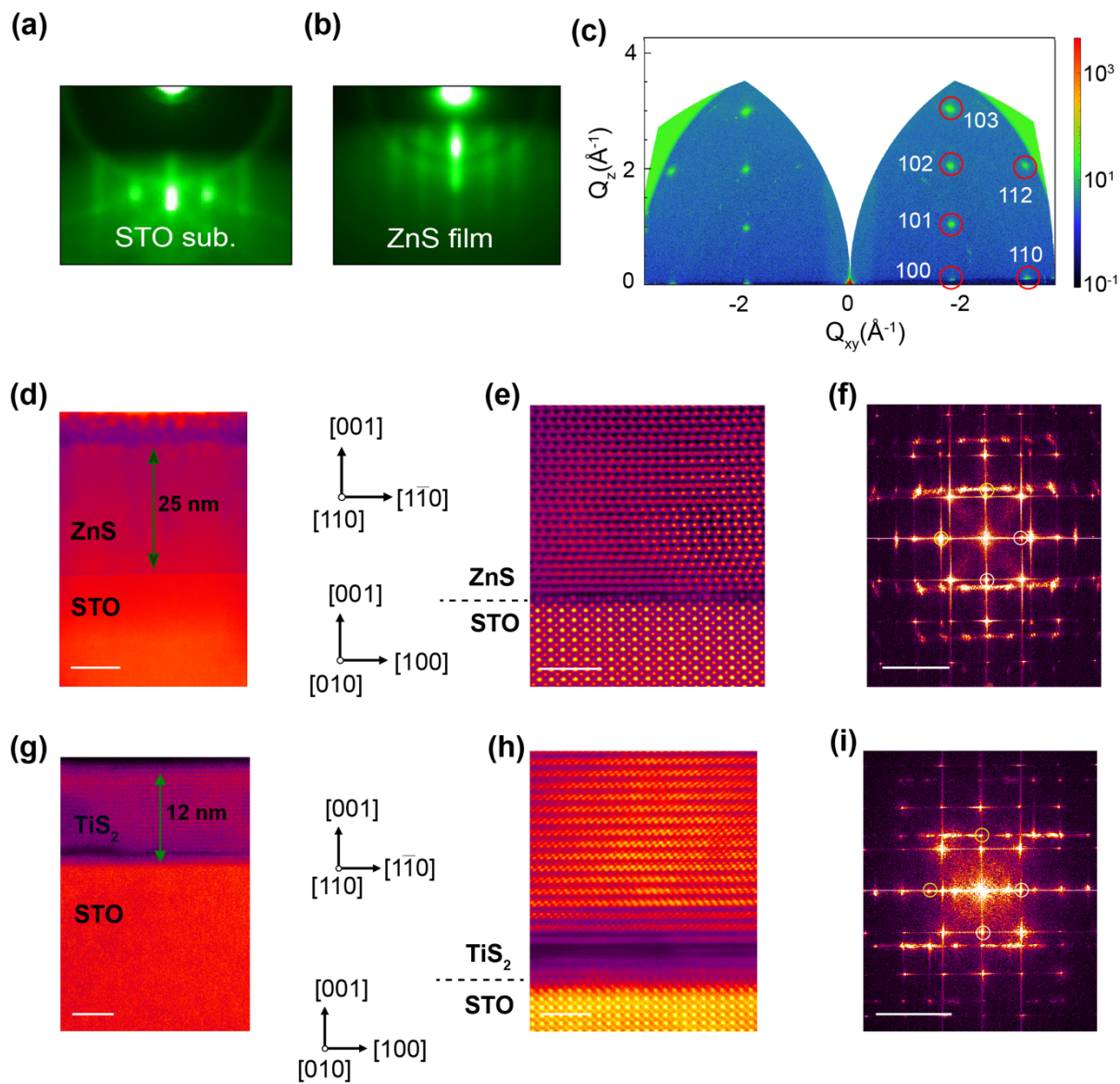

**Fig. 3**. Representative RHEED pattern for **(a)** annealed STO substrate prior to deposition, and **(b)** ZnS thin film after deposition showing streaky pattern. **(c)** GIWAXS pattern for wurtzite ZnS film at an incident angle of 0.5° showing different grain orientations present. Wide field-of-view HAADF image showing **(d)** 25 nm ZnS and **(g)** 12 nm TiS$_2$ films grown on an STO substrate. Atomic resolution HAADF image for **(e)** ZnS/STO and **(h)** TiS$_2$/STO interface indicating the epitaxial orientation of the films with respect to the substrate. FFT patterns for the region in (e) & (h) showing **(f)** ZnS and **(i)** TiS$_2$ 100/001 diffraction spots (yellow circles) along [110] zone axis of STO. Scale bars correspond to 10 nm for d), 5 nm for (g), 2 nm for (e) & (h), and 5 nm$^{-1}$ for (f) & (i).

Fig. 3a shows the RHEED pattern for annealed STO substrate along [100] direction. During ZnS growth, we observed a well-defined specular spot and streaky diffraction pattern in RHEED (Fig. 3b) indicating a highly oriented and smooth in-plane direction. GIWAXS measurements in Fig. 3c showed multiple diffraction spots corresponding to wurtzite structure, suggesting highly oriented grains suggesting the existence of specific in-plane crystallographic orientations in ZnS. To confirm the presence of epitaxy, we further performed STEM on the ZnS thin films and observed a sharp interface between the ZnS film and STO substrate. Fig. 3d and e shows a wide field-of-view and a high resolution HAADF image of a 25 nm ZnS/STO film respectively. The in-plane lattice parameter was measured to be ~3.31 Å and the out-of-plane or growth direction lattice parameter was ~3.13 Å revealing a wurtzite ZnS. The FFT of Fig. 3e is shown in the Fig. 3f where diffraction spots corresponding to ZnS 002, ZnS $1\bar{1}0$ (yellow circles), STO 002 and STO 200 (white circles) were observed, thus confirming the epitaxial relationship as (001) $[1\bar{1}0]$ ZnS // (001) [100] STO. Unlike the BaS films, the ZnS-STO interface was atomically sharp and continuous. We speculate the role of chemistry of the metallic species in the sulfide to influence this intermixing, although lower temperature in the case of ZnS could also play a role. It is also noteworthy that ZnS is a much more stable material compared to BaS, as the later can be easily hydrolyzed in the presence of moisture. This suggests greater degradation possibilities for BaS thin films. However, polycrystalline regions were still observed in ZnS thin films, due to the large lattice mismatch between ZnS and STO. Zn and S atoms were uniformly distributed in the film and showed no diffusion into the substrate as can be seen from the EDS maps in (Supplementary Fig. S8).

Lastly, we grew high-quality textured $TiS_2$ by hybrid PLD at lower temperatures with improved crystallinity. Among the dichalcogenides, early transition metal dichalcogenides containing Ti, Zr, and Hf present the most difficulty due to low vapor pressure of these transition metals and propensity to react with oxygen easily. Hence, we chose to grow $TiS_2$ to demonstrate the advantages of the hybrid PLD approach compared to existing approaches. To the best of our knowledge, high-quality epitaxial growth of $TiS_2$ has not been reported thus far. $TiS_2$ films widely reported in the literature were synthesized by high temperature sulfurization of Ti metal or $TiO_2$, resulting in highly defective polycrystalline films. Thus, hybrid PLD proves to be a promising method to facilitate low temperature growth of high-quality TMDCs by employing organosulfur precursors such as TBDS for efficient sulfurization and thus maintaining the sulfur stoichiometry. Strongly *c*-axis oriented films with 00*l* reflections were grown at 400°C on STO, (Fig. 1b) with a rocking FWHM of ~0.04°. Figure 3g shows a wide field-of-view HAADF image of a 12 nm $TiS_2$ grown at 500°C. In contrast to the other chalcogenides studied in this work, $TiS_2$ has a layered structure where individual layers are bonded together by weak van der Waals forces, which can be clearly seen in the high resolution HAADF images in figure 3h. An out-of-plane lattice parameter ~ 5.7 Å was measured corresponding to the inter-chain distance in $TiS_2$. The $TiS_2$-STO interface looks imprecise, although the layers are clearly distinguishable beyond the first three-unit cells. This could presumably be due to the twisting of layers close to the substrate surface during the growth. However, TiS$_2$ is highly textured as evidenced by the FFT pattern in figure 3i, where diffraction spots of both TiS$_2$ and STO can be clearly seen. The diffraction spots of TiS$_2$ 1$\bar{1}$0 can be seen along STO 200, thus confirming an epitaxial relationship of (001) [1$\bar{1}$0] TiS$_2$ // (001) [100] STO.

It is clear from the above discussion that hybrid PLD is a versatile technique to grow a variety of epitaxial chalcogenides, and TBDS serves as an efficient precursor for sulfur incorporation. In case of binary sulfides, hybrid PLD can be employed to grow epitaxial films at relatively lower temperatures and the films reported here possess narrower rocking curve FWHM than those reported in the literature thus far. However, the texture of chalcogenide perovskite thin films doesn't significantly improve by changing sulfur source to TBDS from Ar-H$_2$S (used in our previous studies). In Fig. 4a, XRD scans are shown for BTS films grown at 700°C and 650°C in TBDS precursor and in Ar-H$_2$S background gas. The intensities of the BTS 110 reflection are comparable for films (same thicknesses) grown at 700°C for both TBDS and Ar-H$_2$S. Films grown by hybrid PLD as well as Ar-H$_2$S[29] at 650°C showed non-trivial but poorer texture compared to those grown at 700°C. At 600°C, the films were most likely amorphous (or nanocrystalline and too weak to be observed in XRD) for both TBDS and Ar-H$_2$S. We speculate that sluggish surface mobility of the adsorbed adatoms is the primary limitation to the growth of textured/epitaxial chalcogenide perovskites, whereas addressing reactivity leads to marginal improvements. However, the surface and interface roughness are significantly higher for films grown at 700°C in Ar-H$_2$S compared to TBDS. XRR curve fits yielded a roughness of about 0.7 nm for the BTS films grown at 700°C in TBDS while those grown in Ar-H$_2$S at 700°C showed a roughness > 2nm. This could be attributed primarily to the sputter damage caused by Ar, and secondarily to the reactivity of H$_2$S causing interdiffusion and rougher films. Meanwhile, at 650°C, the films grown by either method were relatively smoother. This suggests that surface mobility of adatoms need to be enhanced by alternative non-thermal methods to achieve low temperature epitaxial growth with smooth interfaces and high crystalline quality. We speculate that a photo-assisted hybrid PLD approach would serve as a path towards achieving this goal, where an organosulfur precursor would ensure the efficient availability of sulfur, whereas photo-illumination aids in enhancing the surface migration of atoms to realize epitaxial growth at even lower temperatures than achieved in this report.

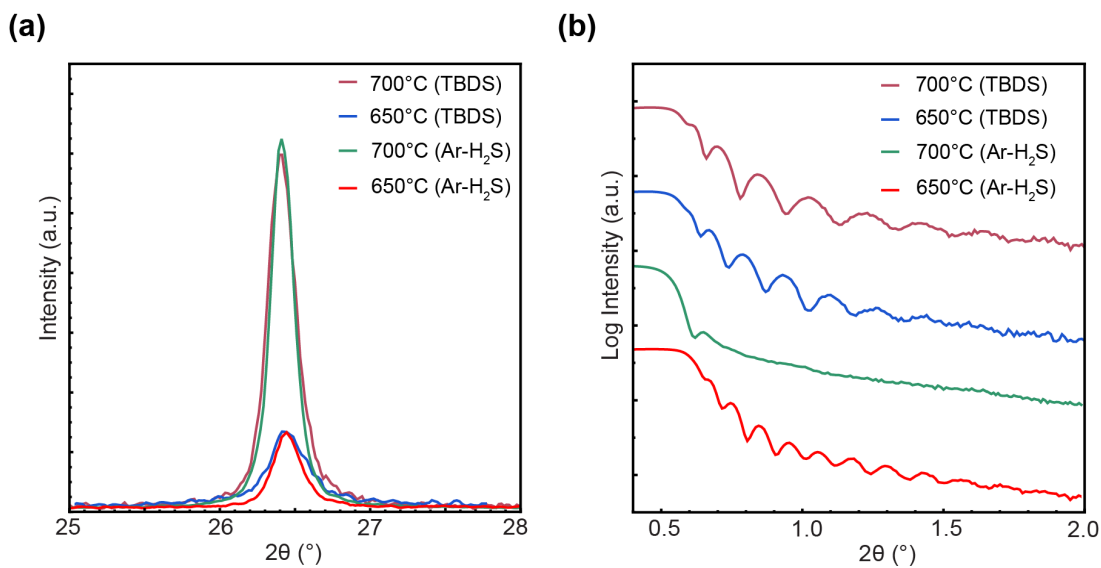

**Fig. 4.** Comparison of **(a)** XRD and **(b)** XRR scans for BTS films grown at 700°C and 650°C by hybrid PLD using TBDS precursor and conventional PLD using Ar-H₂S background gas.

## Conclusion

In this work, we have successfully demonstrated the growth of chalcogenide semiconductors using a novel hybrid PLD approach where an organosulfur precursor is used as an alternative sulfur source to H$_2$S and elemental sulfur. Epitaxial thin films of a broad range of sulfides such as alkaline earth metal chalcogenides, main group chalcogenides, transition metal chalcogenides and chalcogenide perovskites were grown using this method. The films had smooth surfaces and interfaces and showed strong out-of plane textures with narrow rocking curves. We determined epitaxial relationships for all the films grown using a combination of off-axis XRD pole figure scans, high-resolution STEM imaging, GIWAXS measurements and/or *in situ* RHEED analysis. This growth strategy can be extended to the thin film growth of selenides and tellurides as well by the appropriate choice of the organochalcogen precursor. We have demonstrated hybrid PLD as a versatile growth method for chalcogenides, and shows promise to realize epitaxial chalcogenide thin films with high crystallinity and good interface and surface smoothness for electronic and photonic applications in BEOL compatible temperatures. Thus, this novel approach could serve as a pathway to solve some of the growth issues in chalcogenides such as large area growth, cation-chalcogen vapor pressure mismatch and stoichiometric control.

# Methods

**Thin film deposition:** The thin films were grown by PLD using a 248 nm KrF excimer laser in a vacuum chamber specifically designed for the growth of chalcogenides. Single crystal perovskite oxide (Crystec GmbH) substrates such as STO, LAO and sapphire were pretreated by annealing at 1000°C for 3 h in 100 sccm $O_2$ and subsequently cleaned in acetone and IPA prior to deposition. The chamber was evacuated to a base pressure of ~ $10^{-8}$ mbar and then backfilled with high purity argon gas. The substrate was heated up to the growth temperature in vacuum. High purity stoichiometric BZS[28], BTS[29], CaS (Alfa Aesar, 99.9%), SrS (Alfa Aesar, 99.9%) and BaS (Sigma Aldrich, 99.9%), ZnS (Alfa Aesar, 99.99%) and $TiS_2$ powders (synthesized by sulfurization of $TiO_2$ powder (Alfa Aesar, 99.995%) in a carbon disulfide annealing setup at 800°C for 24 hours) were pressed into ¾ inch pellets and densified to > 90% density by room temperature cold isostatic pressing. These dense pellets were used as the targets and were preablated before growth. An organosulfur compound, tert-butyl disulfide (TBDS) (99.999%, Sigma-Aldrich) was used as the sulfur precursor for the PLD growth (Demcon TSST PLD system). Since TBDS is a liquid with relatively high vapor pressure at room temperature, it was introduced by thermal evaporation in a stainless-steel bubbler that was connected through a gas inlet system to the growth chamber. The bubbler was heated to a temperature of 120°C using external heating tapes. The TBDS precursor delivery was controlled using a linear leak valve to achieve total chamber pressures in the range of $10^{-5}$-$10^{-3}$ Torr. No carrier gas was used. The chamber was also equipped with a RHEED system with double differential pumping configuration (TorrRHEED™, Staib) for real-time growth monitoring. The fluence was fixed at 1.0 J/cm$^2$ for all sulfides except BZS (1.8 J/cm$^2$) and the target substrate distance used was 75 mm. The films were cooled postgrowth at a rate of 10°C/min at an Ar partial pressure of 100 mTorr.

**Structural and surface characterization:** The high resolution out-of-plane XRD and off-axis pole figure scans were carried out on a Bruker D8 Advance diffractometer using a Ge (004) two bounce monochromator with Cu Kα1($\lambda$ = 1.5406 Å) radiation at room temperature. XRR measurements were done on the same diffractometer in a parallel beam geometry using a Göbel (parabolic) mirror set up. GIWAXS experiments were conducted in vacuum conditions and at room temperature using a Xenocs Xeuss 3.0 laboratory instrument with an incident wavelength of 1.5406 Å. An incident angle of 0.5° was used for GIWAXS. X-ray scattering data was recorded on a Pilatus 300k area detector at a sample-detector distance of 72 mm. AFM was performed on Bruker Multimode 8 atomic force microscope in peak force tapping mode with a ScanAsyst tip geometry to obtain the surface morphology and roughness.

**Electron Microscopy:** STEM sample preparation was performed using a Thermo Scientific Helios G4 PFIB UXe Dual Beam equipped with an EasyLift manipulator. Standard in situ lift-out technique was used to prepare the TEM lamella. The lift-out region was first coated with the thin e-beam deposited C and W film and followed by a thicker 1.5μm ion-beam deposited W coating to avoid ion-beam damage to the film. The sample was milled and thinned down to about 100 nm using a 30 kV beam and final polishing to remove residual surface beam damage was carried out using a 3 kV ion-beam. STEM experiments were carried out using the probe-corrected Thermo Fisher Scientific Spectra 200 (operated at 200kV) microscope equipped with a fifth-order aberration corrector and a cold field emission electron gun. EDS spectroscopy was carried out using a Dual-X Bruker EDX detectors.

## Supplementary Information

Supplementary information is available for this paper.

## Author Contribution

M.S. and J.R. conceived the idea and designed the experiments. M.S., S.S. and H.C. designed and installed the hybrid PLD setup. M.S., S.S. and C.W. synthesized the PLD targets. M.S. performed the thin film growth and structural and surface characterization. A.A. and Y.T.S. performed the STEM and EDS measurements. All authors discussed the results. M.S. and J.R. wrote the manuscript with input from all other authors.

## Acknowledgements

This work was supported in part by the Army Research Office under Award No. W911NF-19-1-0137, an ARO MURI program with award no. W911NF-21-1-0327, the National Science Foundation of the United States under grant number DMR-2122071. The modification to the growth system to enable hybrid PLD was supported by an Air Force Office of Scientific Research grant no. FA9550-22-1-0117. The authors gratefully acknowledge the use of facilities at the Core Center for Excellence in Nano Imaging at University of Southern California for the results reported in this manuscript. M.S and J.R also acknowledge Prof. Bharat Jalan at University of Minnesota for the helpful conversations and input towards the design of hybrid PLD technique.

Supplementary Information for

# A hybrid pulsed laser deposition approach to grow thin films of chalcogenides

Mythili Surendran[1,2], Shantanu Singh[1], Huandong Chen[1], Claire Wu[1], Amir Avishai[2], Yu-Tsun Shao[1,2], and Jayakanth Ravichandran[1,2,3]*

[1]Mork Family Department of Chemical Engineering and Materials Science, University of Southern California, Los Angeles, California 90089, USA
[2]Core Center for Excellence in Nano Imaging, University of Southern California, Los Angeles, California 90089, USA
[3]Ming Hsieh Department of Electrical and Computer Engineering, University of Southern California, Los Angeles, California 90089, USA

*e-mail: j.ravichandran@usc.edu

## Table of contents

## S1. Hybrid PLD Target Synthesis

Barium titanium sulfide (BaTiS$_3$) and titanium disulfide (TiS$_2$) powders were prepared by sulfurization of corresponding oxide powders in a carbon disulfide annealing setup. Titanium oxide powder (Alfa Aesar, 99.995%)/ barium titanate powder (Alfa Aesar, 99.9%) was loosely packed in an alumina crucible and loaded into a 1" dia tube furnace. The furnace was heated up to 700 °C and 800 °C for BaTiS$_3$ and TiS$_2$ respectively, with a ramp rate of 5 °C/min and held for 24 hours, while Argon was bubbled through CS$_2$ (Alfa Aesar, 99.9 %) with a flow rate of 10 sccm. Afterwards, the furnace was turned off and the sample was allowed to naturally cool down. For barium zirconium sulfide (BaZrS$_3$) powder, stoichiometric quantities of barium sulfide powder (Alfa Aesar 99.7%), zirconium powder (STREM, 99.5%), sulfur pieces (Alfa Aesar 99.999%), and iodine pieces (Alfa Aesar 99.99%) were mixed in a nitrogen-filled glove box and sealed in a quartz tube with around 0.75 mg cm$^{-3}$ iodine, and was held at 960 °C for 150 hours, followed by turning off the furnace. Calcium sulfide powder (Alfa Aesar, 99.9%), strontium sulfide powder (Alfa Aesar, 99.9%), barium sulfide powder (Sigma Aldrich, 99.9%) and zinc sulfide powder (Alfa Aesar, 99.99%) were used as bought. For target preparation, powders were ground finely in a mortar and pestle and pressed into 20 mm pellets using a hydraulic cold press. The pressed pellets were then sintered using an MTI YLJ-100E 100T electric hydraulic press and a 900 MPa CIP (cold isostatic press) die set at ~830 MPa for 5 minutes, to achieve a density of over 90% relative to the respective theoretical densities.

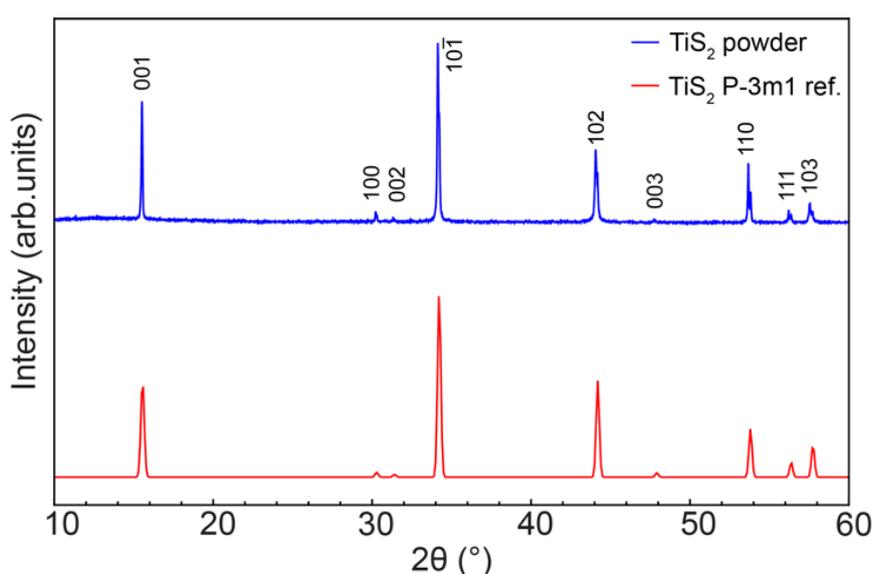

**Figure S1:** Powder XRD scans of TiS$_2$ powder synthesized by CS$_2$ sulfurization of TiO$_2$ powder, along with the TiS$_2$ P$\bar{3}m$1 reference from ICSD.

## S2. Vapor Pressure Table for elements and organosulfur precursors

| Elements | Vapor pressure at 25°C (Torr) | Vapor pressure at 500°C (Torr) |
|---|---|---|
| S | 5.3103 x 10$^{-4}$ | 8.2094 x 10$^4$ |
| Se | 6.1058 x 10$^{-10}$ | 2.2069 x 10$^3$ |
| Te | 1.5237 x 10$^{-13}$ | 36.7955 |
| Ba | 1.2751 x 10$^{-23}$ | 1.3778 x 10$^{-4}$ |
| Sr | 2.4602 x 10$^{-20}$ | 3.7 x 10$^{-2}$ |
| Ca | 3.9833 x 10$^{-23}$ | 4.4048 x 10$^{-4}$ |
| Ti | 4.3041 x 10$^{-73}$ | 4.0992 x 10$^{-22}$ |
| Zr | 1.5549 x 10$^{-95}$ | 6.9854 x 10$^{-31}$ |
| Hf | 4.5649 x 10$^{-99}$ | 2.2905 x 10$^{-32}$ |
| Zn | 1.7561 x 10$^{-14}$ | 1.6480 |

**Table S1.** Vapor pressure of various elements at room temperature and 500°C.

| Organosulfur precursors | Vapor pressure at 25°C (Torr) | Vapor pressure at 100°C (Torr) | Boiling Point (°C) |
|---|---|---|---|
| *Tert*-butyl disulfide (TBDS) | 0.2689 | 17.4034 | 230 |
| *Tert*-butyl sulfide | 3.8717 | 154.9959 | 150 |
| Butyl mercaptan | 181.4571 | 2.0698 x 10$^3$ | 64 |

**Table S2.** Vapor pressure of various organosulfur precursors at room temperature and 100°C along with their boiling points.

## S3. Design and Considerations for Hybrid PLD

The organosulfur precursor was chosen based on the room temperature vapor pressure, decomposition temperature and reaction mechanism. An ideal precursor should have moderately high vapor pressure (> 0.1Torr) at room temperature, and thus the vapors generated at slightly elevated temperatures (~100°C) can be delivered with precise flux control into the vacuum chamber without a carrier gas. The precursor of choice should also have a low decomposition temperature (<300-350°C) to produce sulfur containing species and highly volatile byproducts. In general, a C-S bond is weaker than the H-S bond (in case of H$_2$S) and in organosulfur compounds, it is further weakened when a long hydrocarbon chain is attached to it[1]. Therefore, long chained organosulfides or organodisulfides which can decompose at much lower temperatures than the growth temperatures are promising as sulfur precursors. Among various options, *tert*-butyl disulfide (TBDS) has a suitable vapor pressure at room temperature[2] and undergoes a clean decomposition into highly volatile byproducts (such as isobutene) at low decomposition temperatures of 250-300°C[3, 4]. This ensures a significant reduction in carbon deposition compared to other precursors such as diethyl sulfide and carbon disulfide[5]. Due to the appropriate vapor pressure of TBDS and since no carrier gas is needed, precursor partial pressures as low as $10^{-5}$ Torr are achievable. Moreover, TBDS has been successfully used in the past as an efficient sulfur precursor in ALD and MOCVD growth of transition metal dichalcogenides (TMDC)[2, 4, 6, 7]. Therefore, we chose TBDS to demonstrate the concept of hybrid PLD for chalcogenide thin film growth, although several other precursors such as *tert*-butyl sulfide and butyl mercaptan could be viable and need to be evaluated.

## S4. High resolution XRD of SrS and CaS

CaS ($a$ = 5.683 Å), and SrS ($a$ = 6.007 Å) have a lattice mismatch of about 2.9% and 8.7% with STO substrate in a 45° rotated structure. Regardless of the lattice mismatch, the SrS and CaS thin films were strongly textured on STO substrate with only *00l* reflections observed (**Figure S1**), however, fully relaxed. The rocking curve FWHM of 002 peak of SrS and CaS thin films were 1.21° and 1.32° respectively. Alkaline-earth binary chalcogenide (AeS) films grown on other perovskite substrates like LaAlO$_3$ have a similar texture and rocking curve FWHM while those grown on sapphire showed multiple out-of-plane orientations.

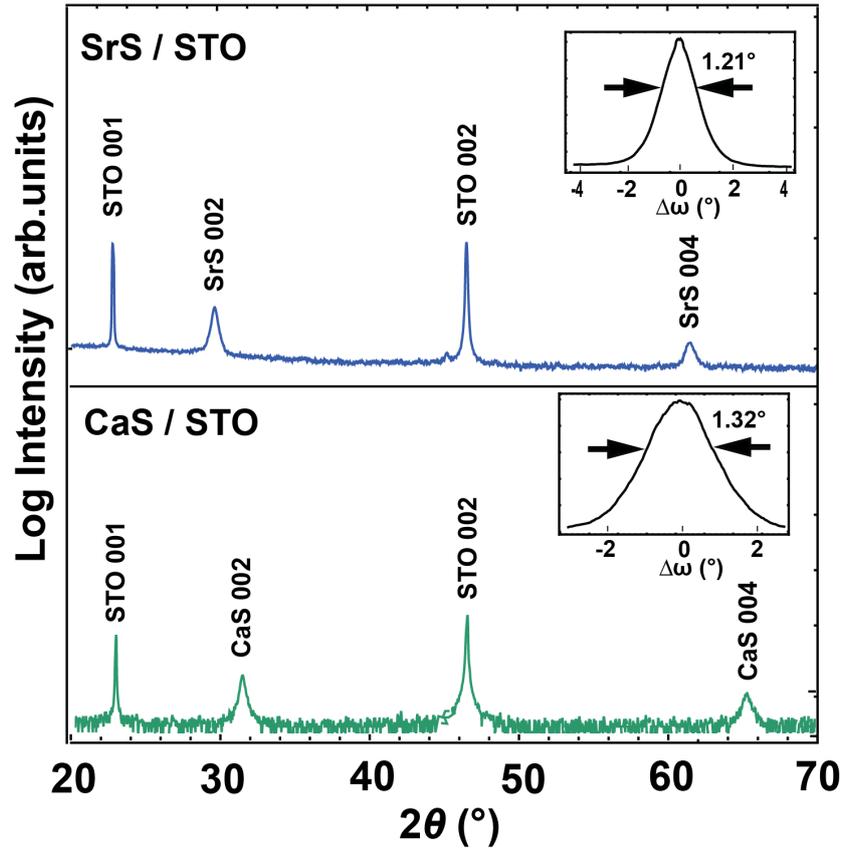

**Figure S2:** High resolution XRD scans of SrS and CaS showing strong out-of-plane texture. The insets show the rocking curves for SrS 002 and CaS 002 peaks.

## S5. Surface and interfacial characterization of alkaline-earth binary chalcogenides

X-ray reflectivity (XRR) measurements were performed to obtain the thicknesses of AeS films. **Fig. S2a** shows XRR curves of 30 nm BaS, 27 nm SrS and 25 nm CaS. The films have smooth surfaces and interfaces as indicated by the slow decay of the reflected X-ray intensity and the presence of Kiessig fringes. Further, AFM topography scans of AeS films confirmed that the film surfaces are smooth with a root mean squared roughness ($R_q$) of less than 1 nm for all the grown films. **Fig. S2b** shows representative AFM images of bare annealed STO substrate and AeS films.

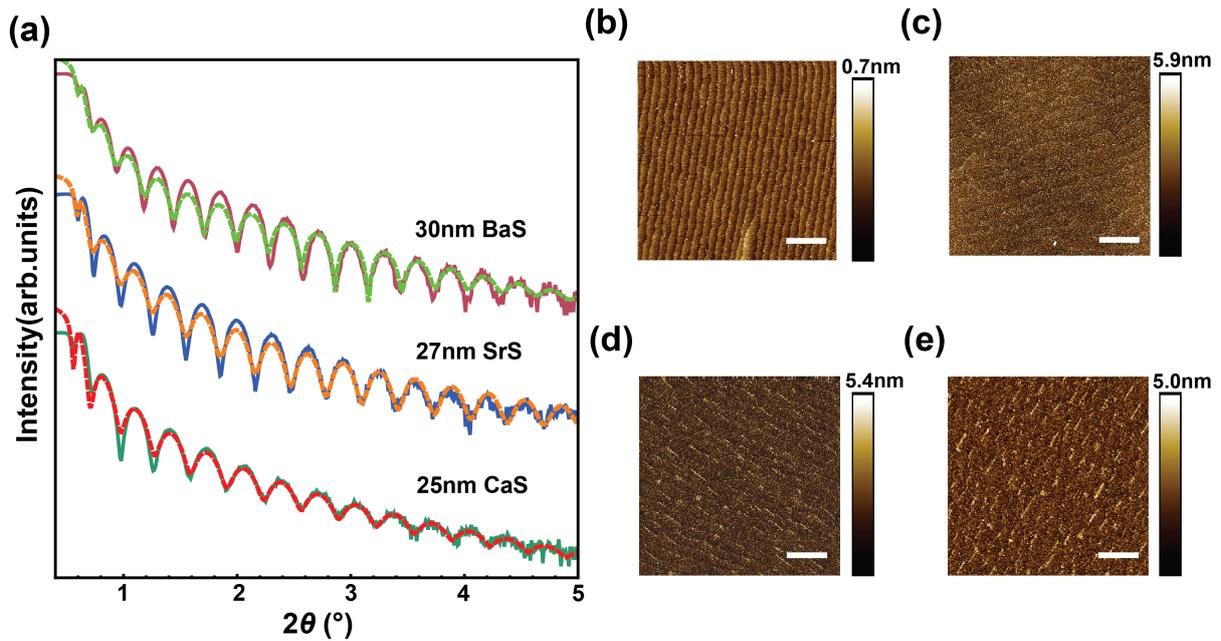

**Figure S3:** a) Measured and simulated XRR curves of AeS films grown on STO substrate along with their respective thicknesses. Representative AFM topography images of (b) bare annealed STO substrate ($R_q$=0.21 nm), (c) 30 nm BaS ($R_q$=0.55 nm), (d) 27 nm SrS ($R_q$= 0.75 nm) and (d) 25 nm CaS ($R_q$= 0.77 nm). Scale bars for (b)-(e) are 1μm.

## S6. STEM images and EDS maps of SrS and CaS

To confirm the epitaxial relationship observed in XRD off-axis $\phi$-scans and to study the SrS (CaS)/STO interface, we performed atomic resolution scanning tunneling electron microscopy (STEM) imaging. **Figure S3** (a) and (c) show a high resolution HAADF image of SrS and CaS respectively, indication an epitaxial relationship of 001) [100] SrS (CaS) // (001) [110] STO. The EDS elemental maps acquired from a wide field of view high-angle annular dark field (HAADF) image suggests a uniform distribution of elements within the bulk of the film (figure S3 (b) and (d)), while the SrS (CaS)/ STO interface was no detectable interdiffusion of elements.

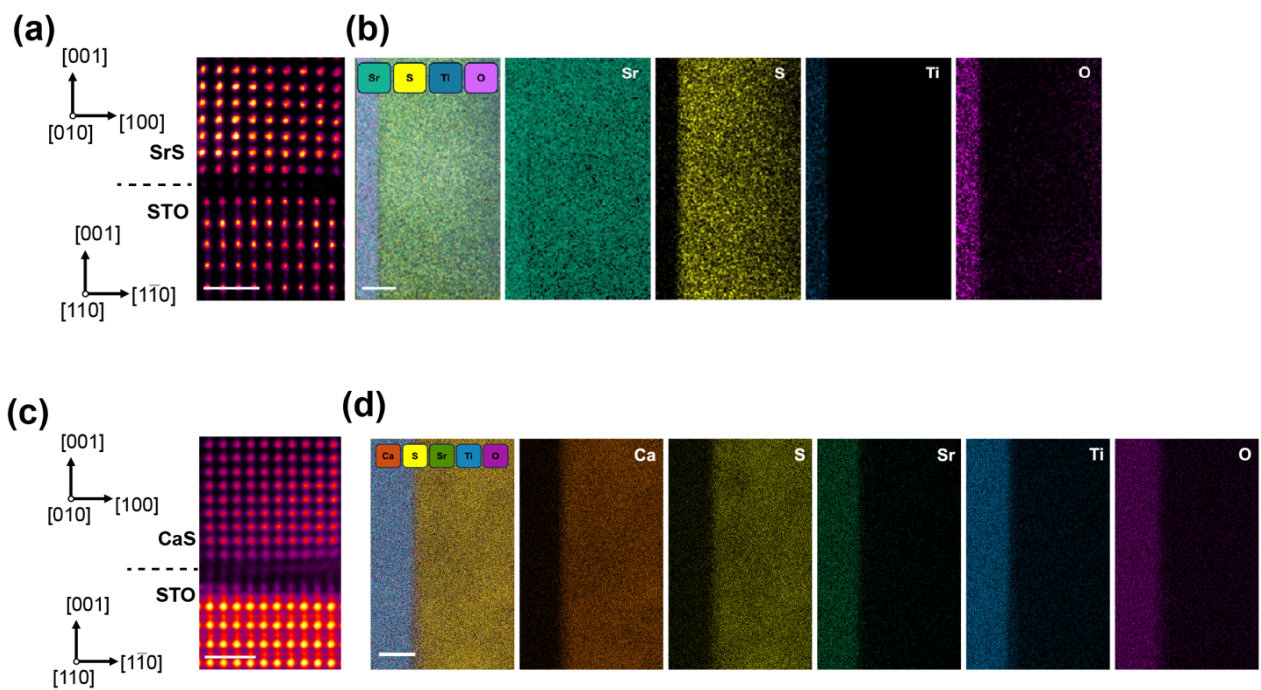

**Figure S4:** Atomic resolution HAADF image for (a) SrS/STO and (c) CaS/STO interface indicating the epitaxial orientation of the film with respect to the substrate. Survey HAADF images of (b) SrS/STO and (d) CaS/STO interface and corresponding elemental maps showing uniform distribution of all elements. Scale bars correspond to 1 nm for (a) and (c) and 10 nm for (b) and (d).

## S7. Structural and surface characterization of ZnS films

Epitaxial ZnS has been extensively grown in the past, however, the structural quality of the films was either inferior due to the growth method itself or carried out at high temperatures. In other cases, high-quality ZnS with rocking curve FWHM = 0.09° (for 2-3 μm films) was reported with photo-assisted MBE and MOCVD techniques, however, the structural characterization was not comprehensive, and illumination was necessary to achieve low temperature epitaxial growth. Wurtzite ZnS films grown at 400°C on STO substrate showed strong out-of-plane texture as shown in XRD scan in figure S4(a). The observation of Laue fringes in the XRD scan indicates a very smooth ZnS/STO interface, irrespective of the dissimilar crystal structure and large lattice mismatch of wurtzite ZnS and cubic STO. This was further confirmed by the extensive Kiessig fringes in the XRR measurements (figure S4(b)). AFM topography image shown in the inset of figure S4(b) also revealed a smooth ZnS film surface with clear step terraces, with a root mean square roughness of about 0.3 nm.

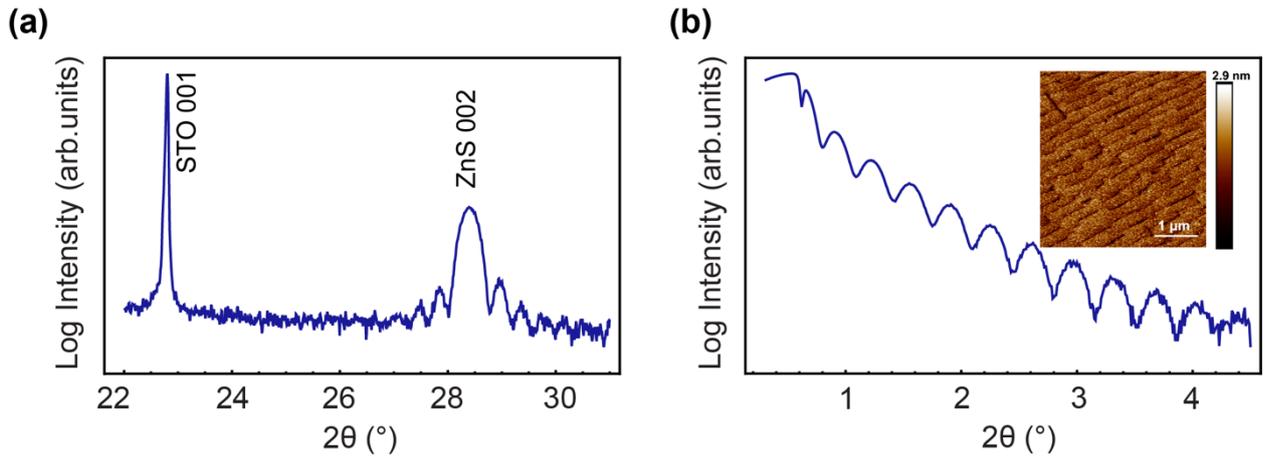

**Figure S5:** (a) High resolution XRD pattern of a representative ZnS film on STO substrate indicating strong out-of-plane texture. (b) XRR curve for a 25 nm ZnS film with inset showing the corresponding AFM image.

However, the XRD off-axis $\phi$-scans of ZnS did not show any direct evidence of epitaxy. This could presumably be due to presence of multiple epitaxial relationships or in-plane polycrystallinity leading to weak diffracted X-ray intensities, limited by the low scattering cross section of ZnS and measurement limits of typical laboratory scale X-ray diffractometers. Therefore, we resorted to a laboratory scale GIWAXS measurement set up with a 2D area detector which can acquire a simultaneous signal collection in both the in-plane and the out-of-plane directions. GIWAXS measurements in Fig. 3c showed multiple diffraction spots corresponding to wurtzite structure, suggesting highly oriented grains suggesting the existence of specific in-plane crystallographic orientations in ZnS.

## S8. Growth temperature and pressure dependence of ZnS growth

Textured ZnS thin films were grown on STO substrates as low as 200°C. The ZnS 002 peak intensity increased as a function of temperature as shown in figure S5, and the best textured films were grown at 400°C with smooth interfaces (figure S4). This is in accordance with the ZnS growths reported in the literature, where epitaxial films were grown at temperatures as low as 175°C by photo-assisted MBE with a narrow rocking curve FWHM of 0.09° for 2-3 μm films. However, all the films grown in this study by hybrid PLD were about 20-25 nm with rocking curve FWHM between 0.02-0.04° (figure 1d). This is impressive and shows the efficiency of this novel method to grow semiconductors compatible with BEOL processing in the CMOS platform.

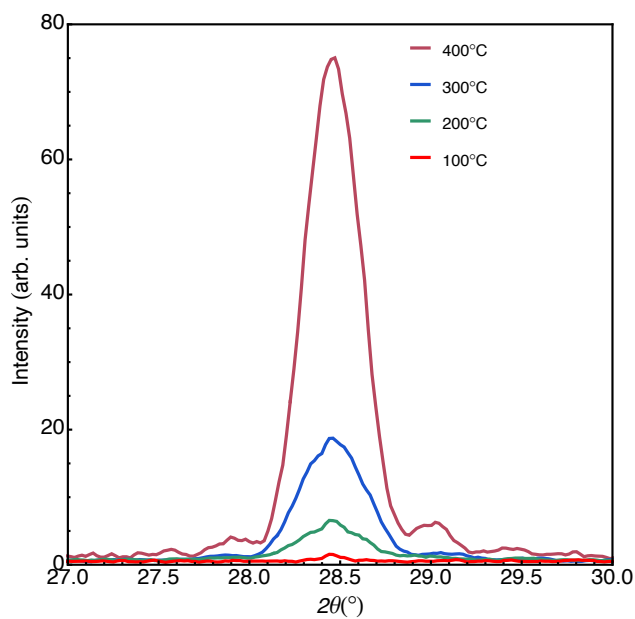

**Figure S6:** High resolution XRD scans showing textured ZnS films grown at temperatures in the range of 200– 400°C. At 100°C, the ZnS 002 peak intensity was very weak, presumably due to nanocrystalline grains.

The ZnS films were grown at 400°C in vacuum as well as in TBDS to show the efficiency of the organosulfur precursor. For ZnS film grown at 400°C in TBDS, the observed out-of-plane lattice parameter was about 3.13 Å corresponding to ZnS 002 interplanar distance as shown in the high resolution XRD scan in figure S6, suggesting a stoichiometric structure and composition. However, ZnS films grown in vacuum had a larger out-of-plane lattice parameter as indicated by a lower $2\theta$ (larger interplanar spacing) in the XRD pattern. This can be attributed to the presence of sulfur vacancies in ZnS films grown in vacuum. At 400°C, the sticking probability of sulfur is lower than at room temperature, hence a sulfur containing background gas is essential during thin film growth to maintain anionic stoichiometry. Here, TBDS serves as an efficient sulfurizing agent and decomposes sufficiently at 400°C to provide the sulfur overpressure during growth. However, in vacuum, the absence of TBDS results in a sulfur deficient ZnS thin film.

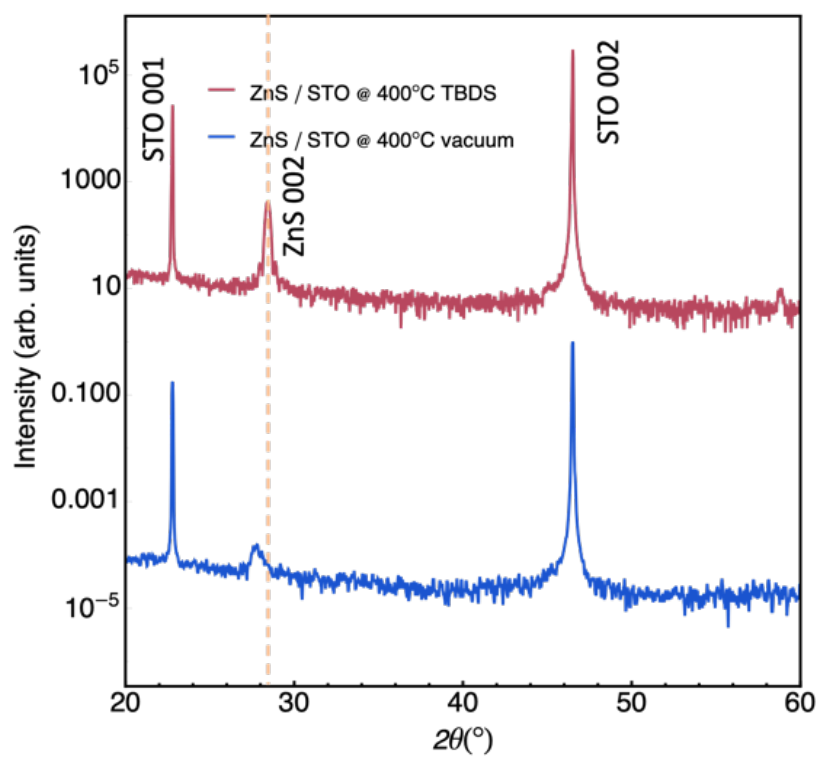

**Figure S7:** High resolution XRD scan of ZnS films grown at 400°C in vacuum and TBDS precursor

## S9. EDS images of ZnS/STO interface

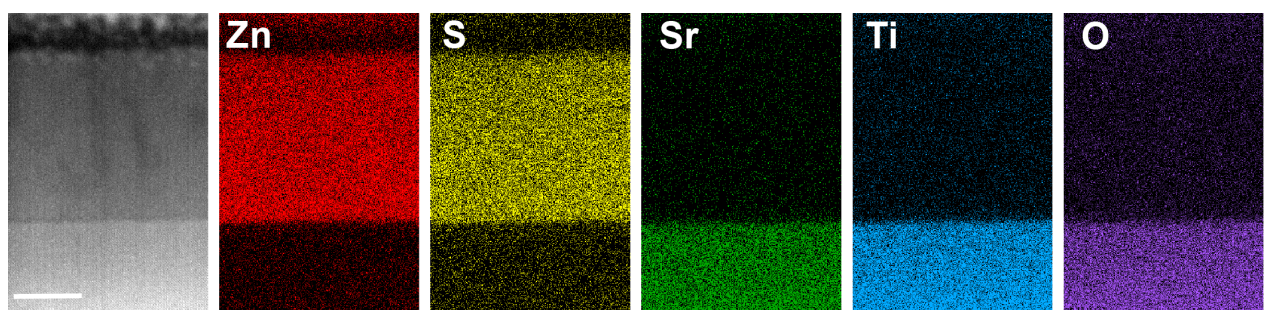

**Figure S8**: Survey HAADF images of ZnS/STO interface and corresponding elemental maps showing uniform distribution of all elements. Scale bar corresponds to 10 nm.

## S10. RHEED pattern of BZS grown on LAO substrate

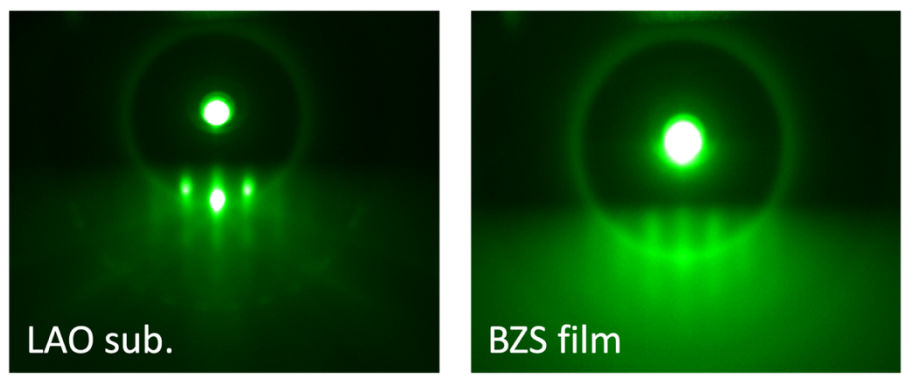

**Figure S9**: RHEED images of LAO substrate at 750°C and BZS film post growth showing streaky pattern.

## S11. Optical images of hybrid PLD grown films

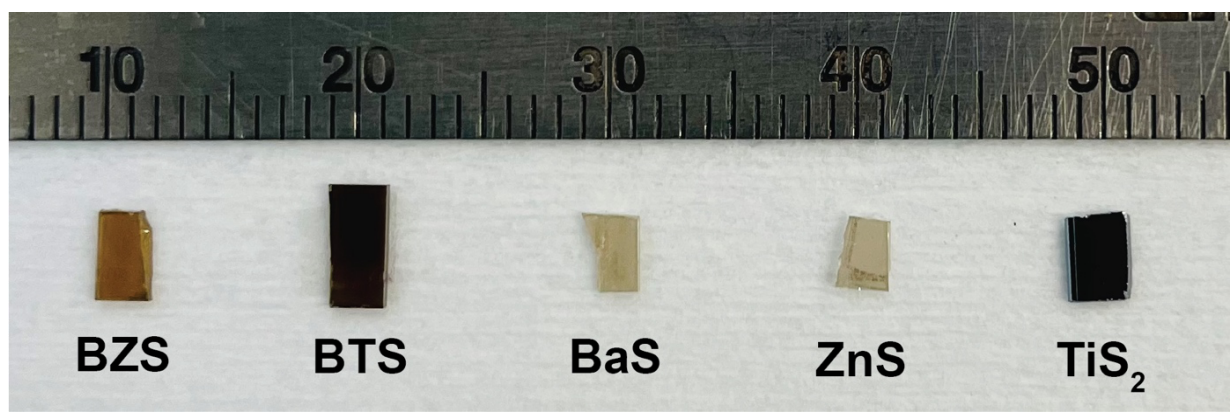

**Figure S10**: Optical images of BZS film on LAO substrate and BTS, BaS, ZnS and TiS$_2$ films on STO substrate. BZS (band gap ~ 1.8 eV), BTS (band gap ~ 0.3 eV), and TiS$_2$ (semimetal) were dark in color, while BaS (band gap ~ 3.8 eV), and ZnS (band gap ~ 3.7 eV) were transparent which is in accordance with their expected band gaps.